\documentclass[aps,prb,twocolumn,groupedaddress,showpacs]{revtex4-1}
\usepackage{amsmath}
\usepackage{graphicx}
\usepackage{amssymb}

\newcommand{\Sprod}[2]{\mathbf{S}_{#1} \cdot \mathbf{S}_{#2}}
\begin{document}


\title{Spin-$1/2$ $J_1-J_2$ Heisenberg antiferromagnet on a square lattice:\\ a plaquette renormalized tensor network study}


\author{Ji-Feng Yu}
\affiliation{Center for Quantum Science and Engineering, National
Taiwan University, No. 1, Sec. 4, Roosevelt Rd., Taipei 106, Taiwan}

\author{Ying-Jer Kao}
\affiliation{Department of Physics, and Center for Advanced Study in
Theoretical Science, National Taiwan University, No. 1, Sec. 4,
Roosevelt Rd., Taipei 106, Taiwan}

\date{\today}

\begin{abstract}
We apply the plaquette renormalization scheme of tensor network
states [Phys. Rev. E \textbf{83}, 056703 (2011)] to study the
spin-1/2 frustrated Heisenberg ${J}_{1}$-${J}_{2}$ model on an
$L\times L$ square lattice with $L$=8,16 and 32. By treating  tensor
elements as variational parameters, we obtain the ground states for
different $J_2/J_1$ values, and investigate staggered
magnetizations, nearest-neighbor spin-spin correlations and
plaquette order parameters. In addition to the well-known N\'{e}el
order and collinear order at low and high ${J}_2/{J}_1$, we observe
a plaquette-like order at ${J}_2/J_1\approx 0.5$. A continuous
transition between the N\'{e}el order and the plaquette-like order
near $J_2^{c_1}\approx 0.40 J_1$ is observed. The collinear order
emerges at ${J}_2^{c_2} \approx 0.62J_1$ through a first-order phase
transition.
\end{abstract}


\maketitle

\section{Introduction}

The search for exotic states in quantum magnets has been the topic
of intensive research for the past decades. An extremely important
question is when the conventional N\'{e}el order is destroyed, what
kind of states can emerge. Frustrated antiferromagnetic spin
systems, where the frustration from  either the lattice geometry, or
the presence of competing interactions, are candidate systems to
study these states.   It is proposed that when the N\'{e}el order is
destroyed by quantum fluctuations, only short-range correlations
will survive, and the system enters a quantum paramagnetic state
which can be described as  a resonant valence bond (RVB)
state.\cite{PAnderson1973} The RVB state can either be a valence
bond solid (VBS) phase, where some of the lattice symmetries are
broken,\cite{NRead1991} or a featureless spin liquid with strong
short-range correlations without any broken spin
symmetry.\cite{FFigueirido1989,LCapriotti2001}  One archetypical
model to study the effect of frustration from competing interactions
is  the antiferromagnetic (AF)
${J}_1$-${J}_2$ Heisenberg model  on a square lattice.\cite{PChandra1988,EDagotto1989,%
FFigueirido1989,HSchulz1992,MZhitomirsky1996,ATrumper1997,RBishop1998,%
LSiurakshina2001,RSingh2003,MMambrini2006,JRichter2010,JReuther2010}
The Hamiltonian is given by,
\begin{equation}
H={J}_1\sum_{\langle ij \rangle}{\mathbf{S}_i\cdot
\mathbf{S}_j}+{J}_2\sum_{\langle\langle ij \rangle\rangle}{\mathbf{S}_i\cdot
\mathbf{S}_j},
\end{equation}
where ${J}_1>0$ and ${J}_2>0$ are the nearest-neighbor (NN) and
next-nearest-neighbor (NNN) couplings, and the sums  $\langle ij
\rangle$ and $\langle\langle ij \rangle\rangle$ run over  NN and NNN
pairs, respectively. Recent interests of this model have been
revived by the discovery of Fe-based superconducting
materials\cite{YKamihara2008} where a weakened AF order can be
described by   this model with $S>1/2$.\cite{TYildirim2008, QSi2008,
FMa2008}

Properties of this model for $S=1/2$ in 2d have been studied
extensively by a variety of methods, such as spin wave
theory,\cite{PChandra1988} exact diagonalization(ED),
\cite{EDagotto1989,HSchulz1992,JRichter2010} series
expansion,\cite{MGelfand1989, VKotov1999, RSingh1999,
OSushkov2001,RSingh2003} large-$N$ expansion,\cite{NRead1991}
functional renormalization group,\cite{JReuther2010} Green's
function method,\cite{LSiurakshina2001} projected entangled pair
states,\cite{VMurg2009} etc. It is generally believed that in the
region ${J}_2/{J}_1\lesssim 0.4$, the ground state (GS) of the model
is the N\'{e}el phase with magnetic long-range order (LRO). In the
region ${J}_2/{J}_1 \gtrsim 0.65$, spins in the GS are ordered at
wave vector $(\pi,0)$ or $(0,\pi)$, showing so-called collinear
magnetic LRO. The GS in intermediate region is proposed to be a
quantum paramagnet without magnetic LRO,  but the properties of this
phase are still under intensive debate. There are several proposals
for the GS, such as  a columnar dimer state,\cite{VMurg2009,
VKotov1999} a plaquette VBS order,\cite{MZhitomirsky1996,
MMambrini2006, LCapriotti2000} or a
spin-liquid.\cite{FFigueirido1989, LCapriotti2001} In the mean time,
precise determination of the phase transition points is also not
conclusive. Earlier series expansion studies\cite{VKotov1999}
estimate the quantum paramagnetic region is between
$0.38\lesssim{J}_2/{J}_1\lesssim 0.62$.     Recent ED
study\cite{JRichter2010} using results of up to $N=40$ to perform
finite-size extrapolation estimates the transition points at
$J_2^{c_1}\simeq 0.35J_1$ and $J_2^{c_2}\simeq 0.66J_1$.
 Meanwhile, studies by combination of random phase approximation and functional
renormalization group find this nonmagnetic phase begins near
$J_2/J_1\approx 0.4\sim0.45$ and ends around $0.66\sim
0.68$.\cite{JReuther2010}

Numerical studies of  frustrated quantum spin systems present
great challenges in  dimensions greater than one. The ED method is hampered by the
 limitation of system size one can simulate. At present, the largest system size on the square lattice that can be simulated is
$N=40$.\cite{ALauchli2005, JRichter2010} Due to the minus sign
problem,\cite{Loh1990} the powerful quantum Monte Carlo (QMC) method
is not applicable to highly frustrated systems. In 1d, the density
matrix renormalization group (DMRG)\cite{SWhite1992} algorithm,
which generates matrix product states (MPS), can reach very high
accuracy  even for  frustrated spin systems; however, direct
extension of the algorithm to higher dimensions remains difficult.
One promising proposal is to generalize the MPS to higher
dimensions, the tensor network states (TNS),\cite{FVerstraete2004,
Nishino1, Vidal1} which can serve as potential candidates for
studying these systems. In the TNSs, the matrices are replaced by
tensors of rank corresponding to the coordination number of the
lattice. On a 2d square lattice, the tensor $T^s_{ijkl}(\sigma_s)$
on site $s$ has four indices, in addition to the physical index,
which in the current case corresponds to the $z$-component
$\sigma_s$ of a spin.

Here, we should mention, according to the TNS representation, the
rank of tensors is chosen according to the coordination number
instead of the interaction pattern. In this way, the area law of
entanglement entropy can be satisfied well if bond dimension D is
big enough, especially when $J_{2}$ not very large.

Contracting over all bond indices gives the wave function
coefficient for a given spin state
$\sigma_1,\ldots,\sigma_N$.\cite{ZGu2008, HJiang2008, plaq} In these
tensor network based methods, one of the major obstacles is the
computational complexity involved in the tensor contraction, then
usually some type of approximation is required to make the
computation manageable. Several schemes have been proposed to
facilitate the contraction of the tensor
networks.\cite{ZGu2008,HJiang2008,plaq,ZXie2009,HZhao2010} In
particular, a contraction scheme based on the plaquette
renormalization with auxiliary tensors
 is proposed to retain the variational nature of the
method, and it is shown that for the transverse Ising model, even
with the smallest possible bond dimension ($D=2$), non-mean-field
results can be obtained.\cite{plaq}

In this paper, we use the TNS with the  plaquette renormalization
scheme to study the $J_1-J_2$ Heisenberg model on a square lattice.
We find that even with a small bond dimension $D=2$, it already
provides a useful way to study the nature of the transition and
estimate the value of the transition points. The rest of this paper
is organized as follows. In the following section, we review the
plaquette renormalization scheme of TNS, and how to apply the scheme
to the current model. Main results will be presented in
Sec.~\ref{sec3}, as well as some discussions. Sec.~\ref{sec4} will
give a brief summary.

\section{\label{sec2}method}

We  investigate the ground state of frustrated Heisenberg
${J}_{1}$-${J}_{2}$ model on a square lattice, using the plaquette
renormalized tensor network\cite{plaq}.  The trial wave function is
written as
\begin{equation}
|\Psi\rangle = \sum_{\{\sigma\}}{tTr(T_1^{\sigma_1} \otimes T_2^{\sigma_2}  \cdots) |\sigma_1\sigma_2\cdots \rangle}, \label{equ2}
\end{equation}
where $tTr$ indicates the tensor trace that all the tensor indices
are summed over.  $T_s$ is rank-4 tensor on site $s$, with bond
dimension $D$ for each rank and $\sigma_s = \uparrow$ or
$\downarrow$ is the physical spin state.

\begin{figure}[tbp]
\begin{center}
\includegraphics[width=8cm,clip]{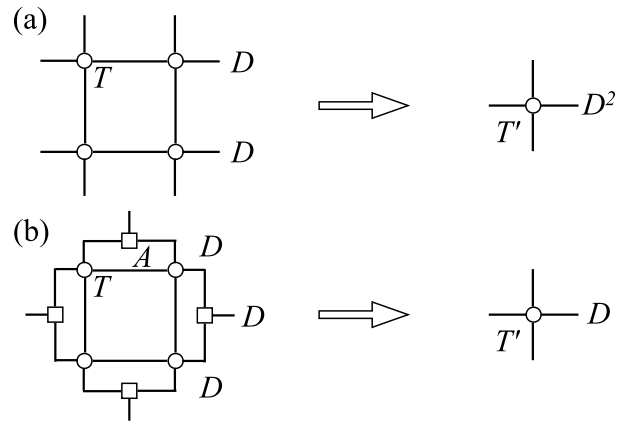}
\caption{\label{fig1}(a) Direct contraction of four connecting
rank-4 tensors $T$ with bond dimensions $D$ results in a new tensor $T'$ with bond dimensions $D^2$ ; (b) Plaquette renormalized tensor contraction via additional auxiliary rank-3 tensors $A$ with bond dimensions $D$. The resulting  tensor $T'$ has the same bond dimension $D$ as the original tensor $T$.}
\end{center}
\end{figure}

Explicit contraction of the  tensor network is computationally
intensive. To keep the computational complexity from growing
exponentially,  auxiliary rank-3  tensors  $A^n_{ijk}$ are added to
each level of the contraction process (Fig.~\ref{fig1}), each
transforms and truncates a pair of indices. A sequence of plaquette
renormalizations, $n=1,2,\ldots$, is carried out and the bond
dimension of each rank is thus kept constant after every plaquette
contraction.\cite{plaq} In order to compute physical expectation
values based on a TNS, one has to contract the tensors of a bra and
ket state over their physical (e.g., spin) indices in addition to
the bond indices of the tensors. Normally, one would first construct
the double tensors by performing the sum over the physical indices,
\begin{equation}
\mathbb{T}^s_{abcd}=
\sum_{\sigma_s,\sigma'_s=\uparrow,\downarrow}T^{s*}_{i_2j_2k_2l_2}(\sigma'_s)T^s_{i_1j_1k_1l_1}(\sigma_s),
\label{decont}
\end{equation}
where the labels $a,b,c,d$ is a suitable combination of the indices
of the bra ($T^{s*}$) and ket ($T^s$) tensors, i.e.,
$a=i_1+D(i_2-1)$, etc.  In the calculation of the matrix element
$\langle \Psi|\hat{O}|\Psi\rangle$ of some operator involving one or
several sites, similar tensors are constructed for the sites at
which operators act weighted with a local expectation value $\langle
\sigma'_{s} | \hat{O}_s |\sigma_s\rangle$. In addition, the
renormalization double tensors can be also formed
\begin{equation}
\mathbb{A}^n_{abc}=A^{n*}_{i_2j_2k_2}A^{n}_{i_1j_1k_1}
\end{equation}
The bond dimension of each rank in the resulting double tensor
becomes $\mathbb{D}=D^2$. This renormalization scheme reduces the
maximum computational complexity\cite{plaq} to $\mathbb{D}^8=D^{16}$
for a double tensor network.

The ground state wave function can be  obtained by optimizing the
elements of tensors $\mathbb{T}, \mathbb{A}$ for the ground state
energy.  Since the plaquette renormalization is introduced at the
wave function level, instead of the constructed double tensor
network, the method remains variational and the final energy will
give a upper bound for the true ground state energy. We optimize the
wave function using the derivative-free Brent's method.\cite{praxis}
Compared to previous methods involving singular value decomposition
(SVD),\cite{HJiang2008, ZGu2008} the environment of a given tensor
is fully taken into account in the current scheme.   However, the
introduction of the  renormalization $A$ tensors at the wave
function level effectively reduces the maximum  support of the
entanglement entropy area law in this tensor network. To reduce the
number of free parameters,  we impose symmetries on the trial wave
function. We use a single plaquette, i.e. $2\times 2 = 4$ sites as a
unit cell (Fig.~\ref{fig1}), wherein tensors $T$ on each site and
auxilliary tensors $A_0$ are assumed to be different. This unit is
translated to generate a $4\times 4 $ unit and another set of
auxilliary tensors $A_1$ are added. This procedure is repeated until
the full lattice is generated.  Finally,  the periodic boundary
condition is applied.\cite{plaq}
\begin{figure}[tbp]
\begin{center}
\centerline{\includegraphics[width=8cm]{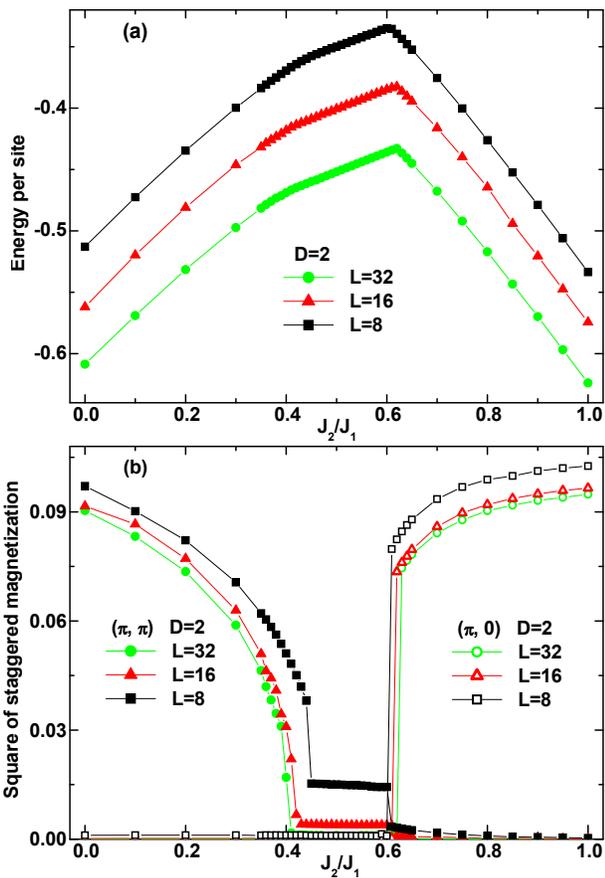}}
\caption{\label{fig2}(Color online) (a) The ground state energy per
site as a function of $J_2/J_1$. The curves for $L=8$ and 16 are
shifted up by $0.05$ and $0.10$ for clarity; (b) The square of
staggered magnetization as a function of ${J}_2/{J}_1$.}
\end{center}
\end{figure}
\section{\label{sec3}results and discussions}
We obtain the ground state wave function by varying the elements in
the tensors $T$ and $A$ with $D=2$, which describes a slightly
entangled state beyond the product (mean-field) state ($D=1$).
Figure~\ref{fig2}(a) shows the ground state energy with  system
sizes $L=8,$ 16, and 32. A clear cusp near $J_2/J_1=0.62$ is
observed,  signaling a first-order phase transition. A continuous
change of the slope is found near ${J}_2/{J}_1 =0.4$, probably
indicating a continuous phase transition there.

To study the details of the magnetic orders and the transition
points, we compute the magnetic structure factor, or the square of
staggered magnetization at wave vector $\mathbf{q}$, defined as
\begin{equation}
 M^2(\mathbf{q}) =\frac{1}{N^2} \sum_{ij} e^{i\mathbf{q}\cdot (\mathbf{r}_i-\mathbf{r}_j)}\left\langle \mathbf{S}_i \cdot \mathbf{S}_j \right\rangle,
\end{equation}
where $\mathbf{r}_i =(x_i, y_i)$, and $\mathbf{q}=(\pi, \pi)$ for
the N\'{e}el order, and $(0, \pi)$ or $(\pi, 0)$ for the collinear
order. $M^2(\mathbf{q})$ tends to the square of the order parameter
in the thermodynamic limit if there is magnetic ordering at wave
vector $\mathbf{q}$, and scales like $1/N$  in a magnetically
disordered phase.

\begin{figure}[tbp]
\begin{center}
\includegraphics[width=8cm,clip]{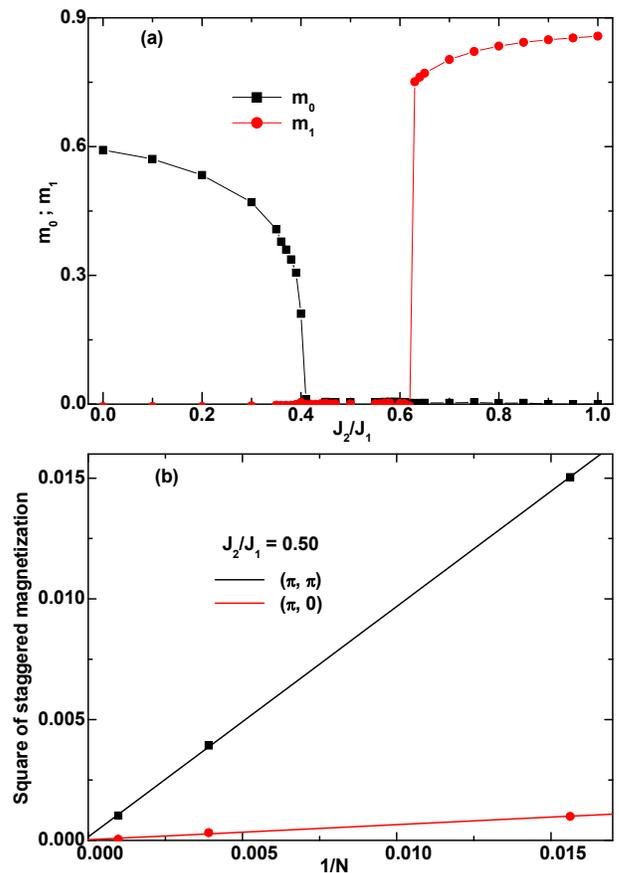}
\caption{\label{fig3} (Color online) (a) Extrapolated order
parameters $m_0$ and $m_1$ as a function of $J_2/J_1$. (b)
Finite-size scaling of $M^2(\pi,\pi)$ and $M^2(\pi,0)$ at
$J_2/J_1=0.5$, where both order parameters $m_0$ and $m_1$ scale to
zero in the thermodynamic limit. }
\end{center}
\end{figure}

Figure~\ref{fig2}(b) shows the results of the square of staggered
magnetizations $M^2(\pi,\pi)$ and $M^2(\pi,0)$. From the small
$J_2/J_1$ side,  the N\'{e}el order is smoothly suppressed as $J_2$
increases, until $J_2/J_1\simeq 0.40$, where a discontinuous jump of
the N\'{e}el order is observed for  $L=8$, and  the jumps become
less pronounced as the system size increases. This strong size
dependence of the jump  is another example that in a finite-size
tensor network state with finite bond dimensions, there exists two
energy minima near the transition,  rendering the transition
first-order at small $N$. For a putative continuous transition,
these two  minima move closer to each other  with increasing $N$ and
the transition becomes continuous at $N\rightarrow\infty$.\cite{CLiu2010}
\begin{figure}[tbp]
\begin{center}
\includegraphics[width=0.45\textwidth]{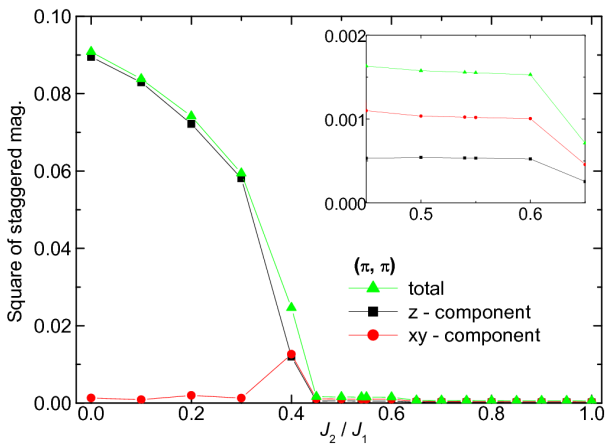}
\caption{ \label{figSymm} (Color online) The $z$(black) and
$xy$(red) components of the square of staggered magnetization and
the sum of the two (green) as a function of ${J}_2/{J}_1$. (Inset)
Same quantities in the regime of ${J}_2/{J}_1 = 0.45\sim 0.65$.}
\end{center}
\end{figure}

From the large $J_2/J_1$ side, the collinear order also decreases
smoothly, until $J_2/J_1 \simeq 0.6$ where a clear first-order
transition occurs. Unlike the previous case, the jumps in
$M^2(\pi,0)$ remain robust upon increasing $N$, strongly suggesting
against a continuous transition here. This transition to the
collinear order is consistent with previous numerical
calculations.\cite{MGelfand1989, VKotov1999, RSingh1999,
OSushkov2001,RSingh2003,JRichter2010}
\begin{figure}[tbp]
\begin{center}
\includegraphics[height=8.5cm,clip,angle=270]{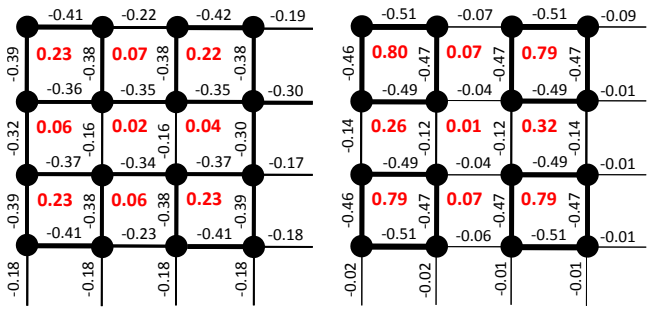}
\caption{\label{fig5} (Color online) The NN spin-spin correlations
$\langle\Sprod{i}{j}\rangle$ (black numbers near bond) and the
plaquette order parameter (red numbers in italic) for ${J}_2/{J}_1=
$ (a) 0.10 and (b) 0.50, with system size $L=32$. We show only one
corner ($4\times 4$) of the entire lattice as the pattern is
repeated periodically.}
\end{center}
\end{figure}

We now use our data from different sizes to extract the order
parameters in the thermodynamic limit. This allows us to estimate
the transition points between the N\'{e}el/collinear state and the
non-magnetic (disordered) phase. The finite-size extrapolation rules
for the two-dimensional antiferromagnetic Heisenberg model  are
well-known\cite{ASandvik1997,PHasenfratz1993, HSchulz1996}.
Following Refs.~\onlinecite{HSchulz1996,JRichter2010}, we define
the N\'{e}el order parameter  as $m_0=2\lim_{N\rightarrow \infty}
M(\pi,\pi)$. This normalization is chosen so that $m_0=1$ in a
perfect N\'{e}el state. The finite-size behavior of $M^2(\pi,\pi)$
is given by,\cite{HSchulz1996,JRichter2010}
\begin{equation}
M^2(\pi,\pi)=\frac{m_0^2}{4}\left(1+\frac{0.62075\ c}{\rho L}+\cdots\right)
\label{eq:m0}
\end{equation}
where $c$ is the spin-wave velocity and $\rho$ is the spin
stiffness. The order parameter for the collinear order is defined as
$m_1=\sqrt{8}\lim_{N\rightarrow \infty} M(\pi,0)$. The finite-size
behavior of $M(\pi,0)$ is given by,\cite{HSchulz1996,JRichter2010}
\begin{equation}
M^2(\pi,0)=\frac{1}{8}m_1^2+\frac{\rm const.}{L}+\cdots
\end{equation}
The extra $1/2$ factor comes from the fact that the ground state has
an extra two-fold degeneracy $\mathbf{q}=(\pi,0), (0,\pi)$, and this
symmetry is broken in the thermodynamic limit. Figure~\ref{fig3}(a)
shows the extrapolated results for $m_0$ and $m_1$ as a function of
$J_2/J_1$. We find that the GS near $J_2/J_1=0.5$ is magnetically
disordered, i.e., both $m_0$ and $m_1$ vanish. Figure~\ref{fig3}(b)
shows the finite-size scaling of $M^2(\pi,\pi)$ and $M^2(\pi,0)$ at
$J_2/J_1=0.5$, which both shows a $1/N$ scaling with the zero
intercept as $N\rightarrow \infty$. The transition points are
estimated to be $J_2^{c_1}=0.40J_1$ and $J_2^{c_2}=0.62J_1$,
consistent with  estimates from   series expansion
\cite{MGelfand1989, VKotov1999,
RSingh1999,OSushkov2001,RSingh2003,JSirker2006} where
$J_2^{c_1}\approx 0.38J_1$ and $J_2^{c_2}\approx 0.62J_1$, and
slightly different from ED results $J_2^{c_1}\approx 0.35J_1$ and
$J_2^{c_2}\approx 0.66J_1$. \cite{JRichter2010} Near $J_2^{c_1}$, we
fit the N\'{e}el order parameter $m_0$ to a power law $m_0\sim
(J_2-J_2^{c_1})^\beta$, and an asymptotic mean-field behavior
consistent with $\beta=1/2$ is also observed.\cite{CLiu2010} For
$J_2=0$, we obtain $m_0 = 0.592$ which is slightly lower than the
best estimate from the quantum Monte Carlo ($m_0 =
0.6140$).\cite{ASandvik2010}  Although it is also possible to
extract $c$ and $\rho$ from our data based on Eq.~(\ref{eq:m0}), it
is argued that determination of these quantities by fitting the
prefactors of the leading finite-size corrections ($O(1/L)$) can not
reach the same accuracy as the magnetic order
parameters.\cite{JRichter2010}

Analogous to how mean-field theory produces symmetry-broken states,
this method can produce solutions which break spin-rotation symmetry
on a finite lattice.\cite{ASandvik2010,CLiu2010} We examine the
spin-rotation symmetry of the ground state, with the focus in the
nonmagnetic phase. Figure~\ref{figSymm} shows $z$ and $xy$
components of the square of staggered magnetization at $q=(\pi,\pi)$
for $L = 32$, defined as
\begin{gather*}
M^2_z(\pi,\pi)=\frac{1}{N^2}\sum_{ij}e^{i\pi[(x_i-x_j)+(y_i-y_j)]}\left\langle S_i^z S_j^z \right\rangle,\\
M^2_{xy}(\pi,\pi)=\frac{1}{N^2}\sum_{ij}e^{i\pi[(x_i-x_j)+(y_i-y_j)]}\left\langle S_i^x S_j^x + S_i^y S_j^y \right\rangle.
\end{gather*}
For reference, the sum of the two is also included. In the N\'{e}el
phase, the spin-rotational symmetry is clearly
broken.\cite{ASandvik2010} Increasing $J_{2}$ through a phase
transition to the strongly frustrated regime (i.e.,
$0.45\lesssim{J}_2/{J}_1\lesssim 0.60$), the spin-rotation symmetry
is restored with $M_z^2 = \frac{1}{2}M_{xy}^2 = \frac{1}{3}M^2$, as
expected.

In order to clarify the possible new phase in the highly frustrated
region around $J_2/J_1=0.5$, we calculate the nearest-neighbor
spin-spin correlations for $L=32$. Figure~\ref{fig5} shows the
results for $J_2/J_1=0.10$, which is deep inside the N\'{e}el phase,
and $J_2/J_1=0.50$, which is in the magnetically disordered phase.
The numbers in black near the bond are the NN spin-spin correlation,
and the thickness of the bond is proportional to its magnitude. For
$J_2/J_1=0.50$ [Fig.~\ref{fig5}(b)], the NN spin-spin correlations
within a single plaquette are much stronger than those between
plaquettes. On the other hand,  deep inside the N\'{e}el phase
${J}_2/{J}_1=0.10$ [Fig.~\ref{fig5}(a)], the NN spin-spin
correlations shows a more uniform pattern, although weaker
correlations are present in some bonds between plaquettes. Overall,
it is clear that the correlations inside a $2\times 2$ plaquette
become stronger upon increasing $J_2/J_1$, which indicates a
possible plaquette order in the magnetically disordered phase.

We also investigate the plaquette order parameter, which
distinguishes clearly a N\'eel ordered phase from a plaquette order,
defined as\cite{VMurg2009}
\begin{eqnarray}
Q_{\alpha\beta\gamma\delta} &=&\textstyle{\frac{1}{2}}(P_{\alpha\beta\gamma\delta} + P^{-1}_{\alpha\beta\gamma\delta}) = 2\big[(\Sprod{\alpha}{\beta})(\Sprod{\gamma}{\delta})\nonumber\\
& + &   (\Sprod{\alpha}{\delta}) (\Sprod{\beta}{\gamma})-(\Sprod{\alpha}{\gamma})(\Sprod{\beta}{\delta})\big]\nonumber\\
&  + & \textstyle{\frac{1}{2}} (\Sprod{\alpha}{\beta} + \Sprod{\gamma}{\delta}+\Sprod{\alpha}{\delta}+\Sprod{\beta}{\gamma}) \nonumber\\
&  + &\textstyle{\frac{1}{2}}(\Sprod{\alpha}{\gamma} + \Sprod{\beta}{\delta} +
\textstyle{\frac{1}{4} }) .
\end{eqnarray}
The results of the plaquette order parameter are shown also in
Figs.~\ref{fig5} (numbers in red italic) for $J_2/{J}_1 = 0.10$  and
$0.50$. In the most frustrated region, we observe  signature of the
plaquette order. For $J_2/J_1=0.50$, the plaquette order parameter
is much stronger within a plaquette, consistent with observation
from the spin-spin correlations. This order parameter is small in
N\'eel phase ($J_2/J_1=0.10$), although some traces of the plaquette
order is still present. This might be due to the inherent structure
of the renormalization scheme, which explicitly breaks the
translational invariance, or possibly the plaquette correlations
already start to build up in this regime. It remains to further
explore whether this plaquette order is favored due to our
renormalization scheme. The plaquette renormalization scheme reduces
the amount of  entanglement  support between plaquettes by a factor
of $D$ compared with the exact contraction. This may bias toward
those correlations compatible with the plaquette structure.

\section{\label{sec4} Conclusion}
We use the plaquette renormalization scheme to study spin-1/2
frustrated Heisenberg ${J}_1$-${J}_2$ model on a square lattice with
different sizes of $L=8, 16$, and 32.  Using the smallest possible
bond dimension $D=2$ for the underlying tensors, we are already able
to obtain results beyond the mean-field theory. Since our method is
variational, and the calculations are done on finite lattices, we
are able to perform finite-size scaling to extrapolate the order
parameters in the thermodynamic limit. We observe signatures of a
continuous transition at $J_2^{c_1}\simeq0.40J_1$, and  a
first-order phase transition at $J_2^{c_2}\simeq0.62 J_1$,
consistent with previous numerical
calculations.\cite{JRichter2010,VKotov1999} Our calculations on the
NN spin-spin correlation and the plaquette order parameter indicates
a possible plaquette VBS order for $J_2^{c_1}< J_2< J_2^{c_2}$.  The
effects of the plaquette renormalization scheme and the bond
dimension $D$ dependence of the physical observables require further
studies and will be presented in a future work.\cite{JYu2011, unpub}

\section{acknowledgements}
We thank A. Sandvik for useful conversation and collaboration on
related work. We are grateful to National Center for
High-Performance Computing Computer and Information Networking
Center, NTU for the support of high-performance computing
facilities. This work was partly supported by the National Science
Council in Taiwan through Grants No. 100-2112-M-002 -013 -MY3,
100-2120-M-002-00 (Y.J.K.), and by NTU Grant numbers 99R0066-65 and
99R0066-68 (J.F.Y., Y.J.K.). Travel support from  National Center
for Theoretical Sciences is also acknowledged.

\emph{note added.}- After submitting this manuscript, we recently
learned of the DMRG work by Jiang \textit{et al.}\cite{HJiang2011}
and the tensor product state approach by Wang \textit{et
al.}\cite{LWang2011} on the same model, which argue that the ground
state in the nonmagnetic regime near $J_2/J_1 \sim 0.5$ could be a
$Z_2$ spin liquid.

\bibliography{J1J2}

\end{document}